\documentclass[aps,prb,amsmath,amssymb,amsfonts,twocolumn,nofootinbib,showpacs]{revtex4-1}
\usepackage{graphicx}% Include figure files
\usepackage{dcolumn}% Align table columns on decimal point
\usepackage{bm,bbm}% bold math
\usepackage{amsmath}
\usepackage{amssymb}
\usepackage{color,hyperref}

\usepackage{soul}
\usepackage[utf8]{inputenc}
\newcommand{\sx}[1]{\sigma^x_{#1}}
\newcommand{\sy}[1]{\sigma^y_{#1}}
\newcommand{\sz}[1]{\sigma^z_{#1}}

\newcommand{\ket}[1]{|{#1}\rangle}
\newcommand{\bra}[1]{\langle {#1}|}
\newcommand{\Li}{\mathcal{L}}
\newcommand{\pp}{\partial}
\newcommand{\D}{\mathcal{D}}
\newcommand{\mc}[1]{\mathcal{#1}}

\newcommand{\up}{\uparrow}
\newcommand{\down}{\downarrow}

\newcommand{\tr}{\mathrm{tr}}

\newcommand{\be}{\begin{equation}}
\newcommand{\ee}{\end{equation}}

\newcommand{\bea}{\begin{eqnarray}}
\newcommand{\eea}{\end{eqnarray}}

\def\Heff{H_{\rm eff}}
\def\opS{S^{\sharp}}
\def\ave#1{\langle #1 \rangle}
\def\etal#1{#1}

\begin{document}
\title{
Influence of dephasing on many-body localization}
\author{Mariya V. Medvedyeva, Toma\v z Prosen, and Marko \v Znidari\v c}
\affiliation{Physics Department, Faculty of Mathematics and Physics, University of Ljubljana, SI-1000 Ljubljana, Slovenia}

\begin{abstract}
We study the effects of dephasing noise on a prototypical many-body localized system -- the $XXZ$ spin $1/2$ chain with a disordered magnetic field. At times longer than the inverse dephasing strength the dynamics of the system is described by a probabilistic Markov process on the space of diagonal density matrices, while all off-diagonal elements of the density matrix decay to zero. The generator of the Markovian process is a bond-disordered spin chain. The scaling variable is identified, and independence of relaxation on the interaction strength is demonstrated. We show that purity and von Neumann entropy are extensive, showing no signatures of localization, while the operator space entanglement entropy exhibits a logarithmic growth with time until the final saturation corresponding to localization breakdown, suggesting a many-body localized dynamics of the effective Markov process.
\end{abstract}


\maketitle

\section{Introduction}

Disorder in a non-interacting system in one dimension leads to unavoidable localization of the electron wave-functions~\cite{Anderson58} due to interference, called single particle localization. Naively, interactions between particles could lead to delocalization, and indeed, at small disorder strength the wave functions overlap significantly, leading to thermalization in finite time~\cite{ETH}. For larger disorder strength though the localization prevails -- the so-called many-body localization (MBL) -- which has been shown to be stable for small interactions~\cite{Gornyi05,BAA06}. The MBL transition can be observed in level spacing statistics~\cite{Vadim07} as well as in dynamic quantities~\cite{dynamics}, like a slow logarithmic growth of the entanglement entropy with time~\cite{MBL08,Moore12}, which is also able to distinguish the MBL and the simpler single particle localized phase. Such MBL behavior can be attributed to the presence of the (quasi)local integrals of motion~\cite{Huse13,Serbyn,intNum,Imbrie14} that guarantee a presence of memory effects, in other words, an MBL system is not ergodic~\cite{Pal10}. For other interesting properties of MBL systems and a more extensive list of references see reviews~\cite{Rahul15,Altman15}.

It is important to understand if an MBL phase persists in the presence of an inevitable coupling to external degrees of freedom, for instance, in experiments~\cite{Schreiber15,Monroe15,Bordia15} probing MBL, that managed to demonstrate non-ergodicity in the presence of controlled interaction and disorder~\cite{Schreiber15}. The influence of an external coupling has been experimentally studied~\cite{Bordia15,talkBloch} with theoretical understanding
though still lacking. It is known for instance that Hamiltonian baths will lead to a broadening of spectral functions~\cite{Rahul14,Sonika15}. Similarly, coupling to another clean~\cite{Bauer15} or disordered~\cite{Bordia15} system can lead to delocalization. We will in particular focus on an experimentaly relevant (and measured~\cite{talkBloch}) dephasing type of dissipation. Dephasing in general arises due to a coupling that preserves local magnetization (or the particle number, in the fermion language, natural for optical lattice experiments). While such coupling can not change the occupation of local sites, it can cause a phase difference between occupational states. This can in turn result in the decay of off-diagonal density matrix elements, leading to a phase
damping, also called dephasing. Alternatively, a Lindblad equation with dissipation of the dephasing type can be derived when there is a fluctuating external field~\cite{Gardiner}, or when one performs a continuous measurement of a site occupation~\cite{Breuer}. The latter is in fact the case in optical lattice experiments~\cite{talkBloch} where off-resonant photon scattering effectively ``measures'' the local site occupation, resulting in dephasing~\cite{Sarkar14}. While it has been shown~\cite{NJP10} that dephasing eventually breaks the MBL phase, resulting in diffusion, we shall study how relaxation to this  state proceeds. We note that the same problem has been considered recently in Ref.~\onlinecite{Levi15}, with different scaling on disorder strength predicted. Another, topological perspective on the behaviour of similar disordered systems under open (dissipative) quantum dynamics has been considered in Ref.~\onlinecite{Carmele}.

In this paper we treat the quantum dissipative dynamics by perturbation theory in the inverse disorder strength.
Such an approach, firstly, provides us with a scaling variable and shows that in one-dimension the interactions between particles do not influence the relaxation. Secondly, it reduces the dimension of the Hilbert space for the operators of interest and allows to obtain scaling functions for the time-evolution of observables in the thermodynamic limit. 

\section{The model}

We shall study a prototypical MBL system in the presence of dephasing, described by a Lindblad equation~\cite{Lindblad} $\frac{\pp \rho}{\pp t}= \Li [\rho]$, where
\begin{equation} 
\Li [\rho] = -i [H,\rho]+\D[\rho],\quad \D[\rho] = \sum_j \left(2 l_j \rho l_j^\dagger  - \{ l_j^\dagger l_j,\rho\} \right),
\label{eq:Lin}
\end{equation}
with Lindblad operators $l_j=\sqrt{\frac{\gamma_j}{2}}\sz{j}$. For the derivation see books in Ref.\onlinecite{Gardiner,Breuer}, or Ref.~\onlinecite{Sarkar14} for the optical lattice context. In MBL experiments~\cite{talkBloch} the dephasing strength $\gamma_j$ depends on the laser intensity and detuning. The Hamiltonian is the canonical MBL system, namely the disordered $XXZ$ spin chain,
\begin{equation}
H = \sum_{j=1}^{L-1} J_j\left( \sx{j}\sx{j+1} + \sy{j}\sy{j+1}+\Delta \sz{j}\sz{j+1}\right) + \sum_{j=1}^L  h_j\sz{j},
\label{eq:H}
\end{equation}
where $L$ is the number of sites and $\sigma_j^\alpha$ are Pauli matrices. Note that this model after Jordan-Wigner transformation becomes a spinless fermionic model, whose spinfull version is the subject of cold atomic experiments~\cite{Schreiber15,Bordia15}. The magnetic field $h_j$ is random and uniformly distributed in the interval $h_j \in [-2h,2h]$ for some disorder strength $h$. In this notation the transition from the ergodic to the MBL phase happens~\cite{Pal10,Luitz15} around $h=h_{\rm c} \approx 3.7$. We are interested in the influence of the dephasing on the MBL phase, so we focus on $h>h_{\rm c}$. The steady state (determined by the condition $\Li [\rho]=0$)  is an infinite temperature state, $\rho\propto \mathbbm{1}$,  as can be easily seen from the Lindblad equation.

{\em Effective description.--}
Since the critical field $h_{\rm c}$ is large, we shall use perturbation theory in the inverse disorder strength $h$ for an effective theoretical description. For not too short times, $t>{\rm max}_j (1/\gamma_j)$, the dynamics is governed by eigenvalues of $\Li$ that are close to the steady-state eigenvalue $0$, and those eigenvalues will be correctly accounted for by our theory. We split the Liouvillian $\Li$ into a ``large'' $\Li_0$ that contains the diagonal part in the computational basis (terms with $\Delta$, $h_j$ and $\gamma_j$) and $\Li_1$ that contains the hopping. Because eigenvalues of $\Li_0$ are large compared to those of $\Li_1$ the pseudoinverse $\Li_0^{-1}$ is uniformly small, resulting in a well behaved perturbation series. Our discussion here closely follows that in Refs.~\onlinecite{Gaps,Cai:13}, where a similar method has been used for clean systems, see also a similar-spirited approach in the context of Rydberg gases~\cite{Lesanovsky13}. $\Li_0$ has a degenerate eigenvalue $0$ with eigenoperators being projectors $\ket{\psi_n}\bra{\psi_n}$ to all $2^L$ joint eigenstates $\ket{\psi_n}$ of $\sigma^z_j$ ({\em computational basis}). In the first order the degeneracy is not removed (only off-diagonal eigenoperators of $\Li_0$ are affected), while the second order degenerate perturbation theory gives 
\be 
\Li_{\rm eff} = -\mc{P} \Li_1 (1-\mc{P})\Li_0^{-1}(1-\mc{P})\Li_1 \mc{P},
\ee 
where $\mc{P}$ is a projector to the eigenspace of $\Li_0$ with eigenvalue $0$. Because $\Li_{\rm eff}$ is Hermitian, we write its matrix elements as $[\Heff]_{m,n}:=-\tr{[\ket{\psi_m}\bra{\psi_m}\Li_{\rm eff}(\ket{\psi_n}\bra{\psi_n})]}$, resulting in (see Appendix)
\be \label{eq:Heff}
\Heff \approx \sum_j  \frac{ 2J_{j}^2 (\gamma_j+\gamma_{j+1})}{(h_{j+1}-h_j)^2} (1-\vec{\sigma}_j \cdot \vec{\sigma}_{j+1})
\ee 
where we write explicitly only the leading order (quadratic) terms in $1/h$. From now on we set $J_j\equiv J$ and $\gamma_j \equiv \gamma$. As we shall demonstrate, for times much larger than $1/\gamma$, the dynamics of $\rho(t)$ is correctly described by $\Heff$~\cite{footg}. Namely, on a time scale 
$\sim 1/\gamma$ all off-diagonal terms in $\rho(t)$ decay to zero, and what remains for $t> 1/\gamma$ is a process generated by $\Heff$ on the linear space 
spanned by projectors $\ket{\psi_n}\bra{\psi_n}$:
\begin{equation}
\rho(t) \approx \sum_n p_n(t) \ket{\psi_n}\bra{\psi_n}, \quad \vec{p}(t)={\rm e}^{-\Heff\, t} \vec{p}(0),
\label{eq:p}
\end{equation}
where $\vec{p}$ is a vector of probabilities $p_n$. Although the generator is a ``quantum'' matrix, of size $2^L$, it can be considered as a generator of a classical Markov process. In particular, the propagator is a stochastic matrix, $\sum_m [{\rm e}^{-\Heff\, t}]_{m,n}=1$, conserving total probability $\sum_n p_n$. $\Heff$ (\ref{eq:Heff}), yielding evolution~(\ref{eq:p}), is the main result of our work. 

Even without further calculations we can draw several important consequences. Evolution with $\Heff$ (\ref{eq:Heff}) immediately implies that the unique scaling variable for $\rho(t)$ (and all observables) is $\tau=J^2 \gamma t/h^2$. Furthermore, $\Heff$, and with it dynamics for $t>1/\gamma$, does not depend on the interaction $\Delta$. The fact that $\Heff$ is always isotropic, regardless of the interaction $\Delta$ in the original $H$, is due to probability conservation (anisotropic $\Heff$ would not result in a stochastic propagator). There is an interesting duality: $\rho(t)$ is again governed by a disordered Hamiltonian, but this time with the disorder in bonds. Eventhough the distribution of bond strengths in (\ref{eq:Heff}) is singular, after exponentiation large eigenvalues do not contribute to
 ${\rm e}^{-\Heff\, t}$ on times larger than $1/\gamma$ (higher order terms left out from $\Heff$ would also make the strongest bond finite, see Appendix A. It is not known whether  $\Heff$ (\ref{eq:Heff}), re-interpreted as a quantum Hamiltonian, displays MBL~\cite{Martin,Vasseur}. Below we shall demonstrate the validity of the description with $\Heff$, in particular showing that higher order terms in the perturbation series are negligible even for not-so-small $h=4$, that the convergence radius does not shrink with the system size, and calculate interesting physical quantities.

\section{Numerical results}

Looking at the spectrum of $\Li$, see Appendix A (Fig.~5), we observe that there are $\approx 2^L$ real eigenvalues, while most $\sim 4^L$ have nonzero imaginary part. The simplest way to understand such a separation is to consider a perturbation theory in $\gamma$ in the unperturbed operator eigenbasis of $\ket{\psi_m}\bra{\psi_n}$. The eigenvalues/eigenoperators corresponding to $m\ne n$ obtain a non-zero real part (the decay rate) in multiples of $2\gamma$. As a consequence, the off-diagonal elements of $\rho(t)$ decay within times $t<1/\gamma$. For  $m=n$ the degeneracy of unperturbed eigenvalues $0$ is lifted by the second order perturbation theory, and those are the eigenvalues of interest to us and are described by $\Heff$ even for a single disorder realization, see~Appendix A.

\begin{figure}[tb]
\centering \includegraphics[width=3.33in]{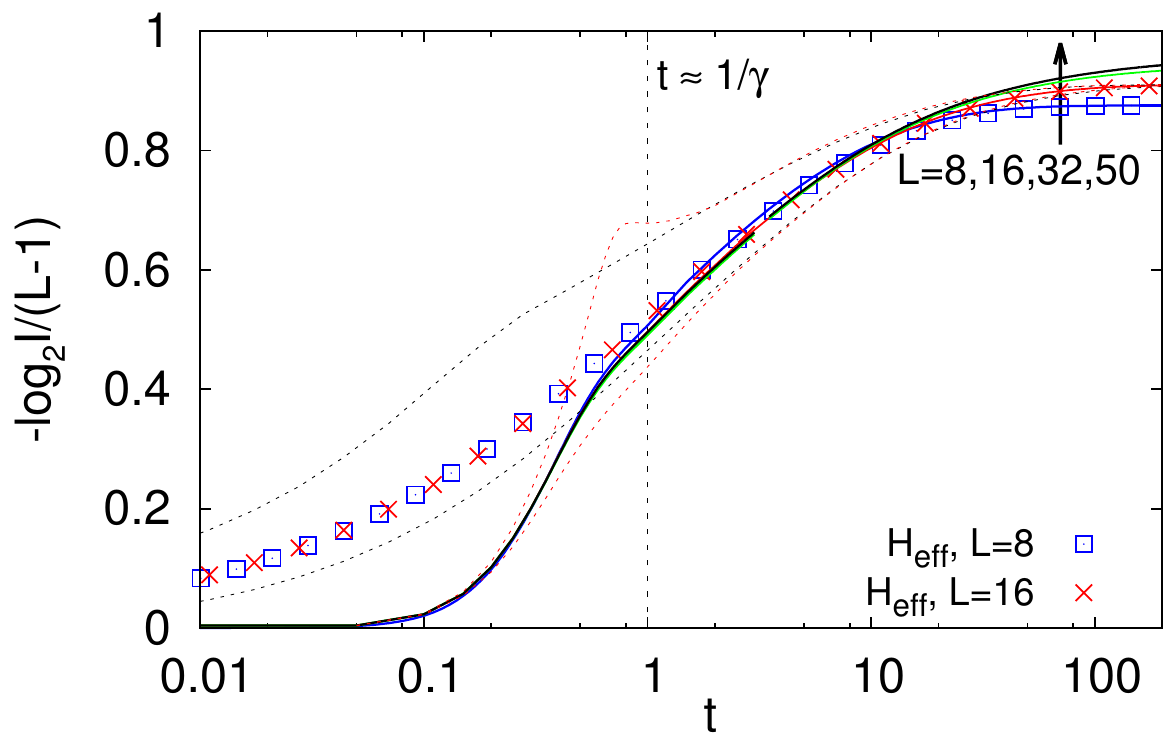}
\vspace{-2mm}
\caption{(Color online) Disorder-averaged purity for exact (full curves, $L=8$ blue, $16$ red, $32$ green, $50$ black) and approximate evolutions with $\Heff$ (squares and croses): for times larger than $t \sim \frac{1}{\gamma}$ the two agree, as well as their fluctuations due to disorder (dotted red/black lines are standard deviations $-\log_2{(I\pm \sigma)}/(L-1)$ for evolution by $\Li$ and $\Heff$, respectively). $\Li$ data for $L=8$ is obtained by exact diagonalization, $L=16,32,50$ with tDMRG (see Ref.~\onlinecite{NJP10} for details of implementation). AF initial state is used, $\gamma=J=1$, $h=4$, $\Delta=0.5$.}
\label{fig:afI}
\end{figure}

\begin{figure}[tb]
\centering \includegraphics[angle=-90,width=3.33in]{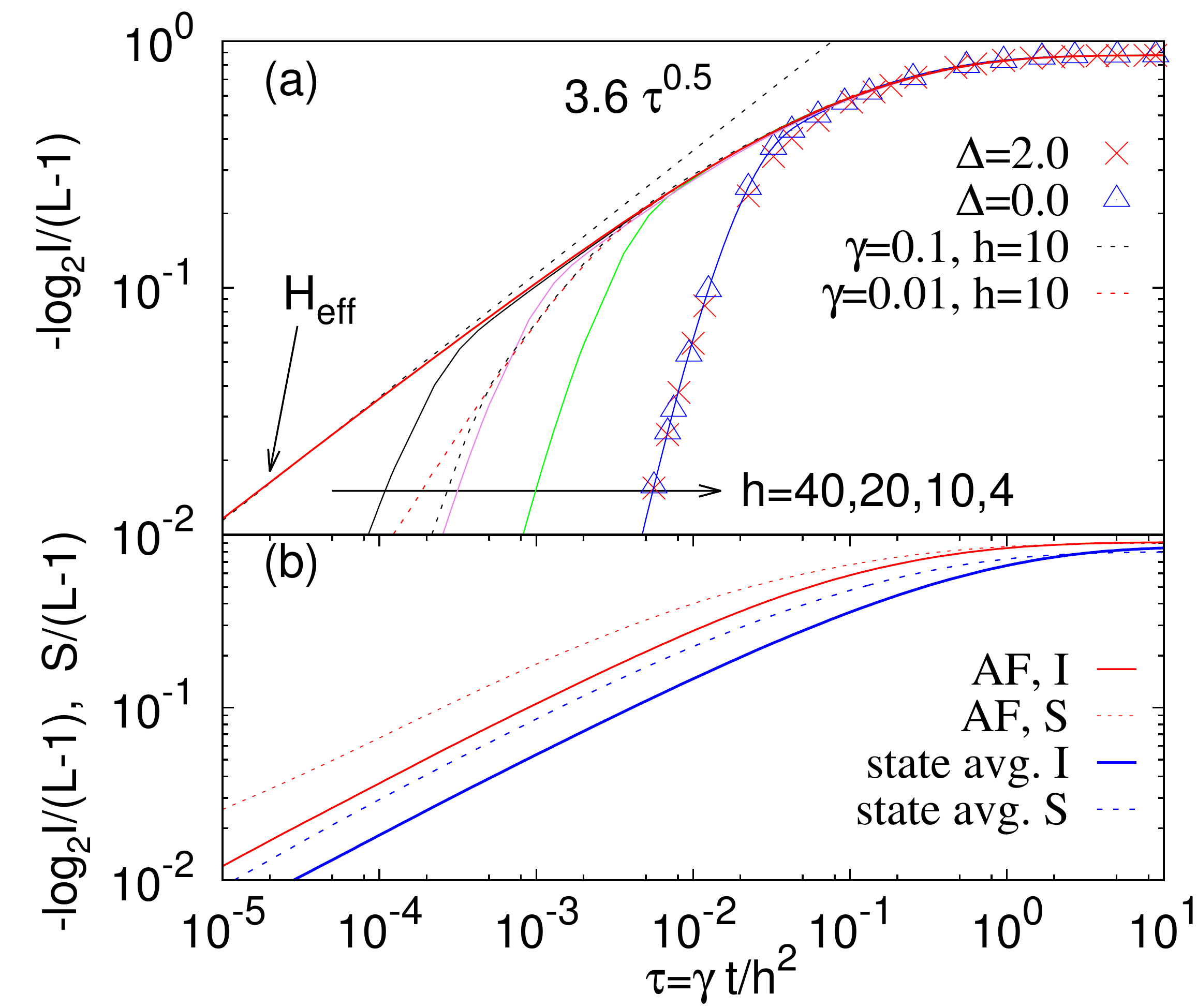}
\caption{(Color online) (a) Purity scaling function (\ref{eq:scalI}) for the AF initial state (full red curve, obtained with $\Heff$ and $L=16$) and exact dynamics with $\Li$ ($L=8$) for different $h$ (full curves), $\gamma$ (two dotted curves) and $\Delta$ (symbols). (b) The scaling function for a disorder-averaged purity $I(t)$ (full curves) behaves essentially the same as a disorder-averaged entropy $S(t)$ (dotted curves). Red curves are for the AF initial state, blue ones for the state-averaged quantities (\ref{eq:avgI}) (all for $\Heff$).
For all curves (in (a) and (b)), all unspecified parameters are $h=4$, $\gamma=J=1$, $\Delta=0.5$.
}
\label{fig:afscal}
\end{figure}

The first quantity that we study is purity $I(t):=\tr \rho^2(t)$. It is simpler for analytical treatment than the von Neumann entropy $S(t):=-\tr \rho(t)\log_2\rho(t)$, and we shall demonstrate that $S(t)$ behaves in essentially the same way as $-\log_2 I(t)$. In our effective diagonal Markov description we have $I(t)=\sum_n p_n^2(t)$, and similarly for $S(t)=-\sum_n p_n(t) \log_2 p_n(t)$. Note that $S(t)$ is not a measure of quantum entanglement of $\rho(t)$ -- the state is
diagonal in computational basis and hence non-entangled for $t>1/\gamma$. For the initial state we first choose the antiferromagnetic (AF) state $\ket{\down\up\down\up\cdots \down\up}$ also used in experiments~\cite{Schreiber15}. In Fig.~\ref{fig:afI} we show a comparison between a disorder-averaged $I(t)$ calculated with the exact evolution with $\Li$ and the one calculated using $\Heff$ (\ref{eq:Heff}). As predicted, for $t>\frac{1}{\gamma}$ the effective description is correct, not just for the average purity, but also for fluctuations, and even for each disorder realization 
separately. We also see that $-\log_2 I$ is an extensive quantity as it is proportional to $L$. For large times, in the crossover regime to the saturation value, there is a subleading small dependence on the system size, whereas for shorter times the asymptotic behavior for $L \to \infty$ is already reached for smaller $L$ (curves for different $L$ overlap for $t \le 50$). Even though our theory is perturbative in $1/h$, already for $h=4$ a good agreement is achieved.
Next, we verify the predicted scaling of the purity, which is
\begin{equation}
-\log_2 I(t)=(L-1) f(\tau),\qquad \tau=\frac{J^2\gamma t}{h^2},
\label{eq:scalI}
\end{equation}
where $f(\tau)$ is a scaling function (that can depend on the initial state). To that end we show in Fig.~\ref{fig:afscal}(a) the scaling function obtained from $\Heff$ as well as results of exact simulation for different $h,\gamma$ and $\Delta$, all collapsing on the scaling function~\cite{foot2} for $t>\frac{1}{\gamma}$. For large $h$ there is therefore no dependence on the interaction $\Delta$ (for the $XX$ chain with disorder and dephasing some exact results are available~\cite{XXdeph}). For small $\tau$, which is reachable for large $h$, the scaling function behaves as $f(\tau) \sim \sqrt{\tau}$, explanation of which is very simple: shortest times are dominated by largest eigenvalues of $\Heff$ due to diverging bond strength (\ref{eq:Heff}) for resonant $h_j$. Considering just the strongest bond (in the spirit of real space renormalization group) giving eigenvalue $4/(h_1-h_2)^2$, results in disorder averaging $\int dh_1 dh_2\exp{(-16 t/(h_1-h_2)^2)}$. Expansion of this 
 (exact for $L=2$) purity results in a $\sqrt{\tau}$ behavior, which together with $L-1$ possible (largest) bonds gives~\cite{foot1} $-\log_2 I \sim (L-1)\sqrt{\tau}+\cdots$.
We also calculated the state-averaged purity $\ave{I(t)}$, with averaging being done over all initial states $\ket{\psi_n}\bra{\psi_n}$, the expression for which 
is a partition function of $\Heff$ at inverse temperature $2t$~(Appendix B).
\begin{equation}
\ave{I(t)}=\frac{1}{2^L}\tr{[{\rm e}^{-2\Heff\, t}]}.
\label{eq:avgI}
\end{equation}
$\ave{I(t)}$ might be more representative than $I(t)$ for the AF initial state, however, as Fig.~\ref{fig:afscal}(b) demonstrates, it behaves in qualitatively the same way as the AF purity. This thermodynamic nature of $I(t)$ and $S(t)$ explains extensivity of $-\log_2 I(t)$ and $S(t)$ for typical initial states, as well as the fact that there are no MBL-like signatures in $S(t)$ or $I(t)$ because thermodynamic expectation values are not able to detect an MBL phase~\cite{Rahul15,Altman15}.

\begin{figure}[tb!]
\begin{center}
\includegraphics[width=\linewidth]{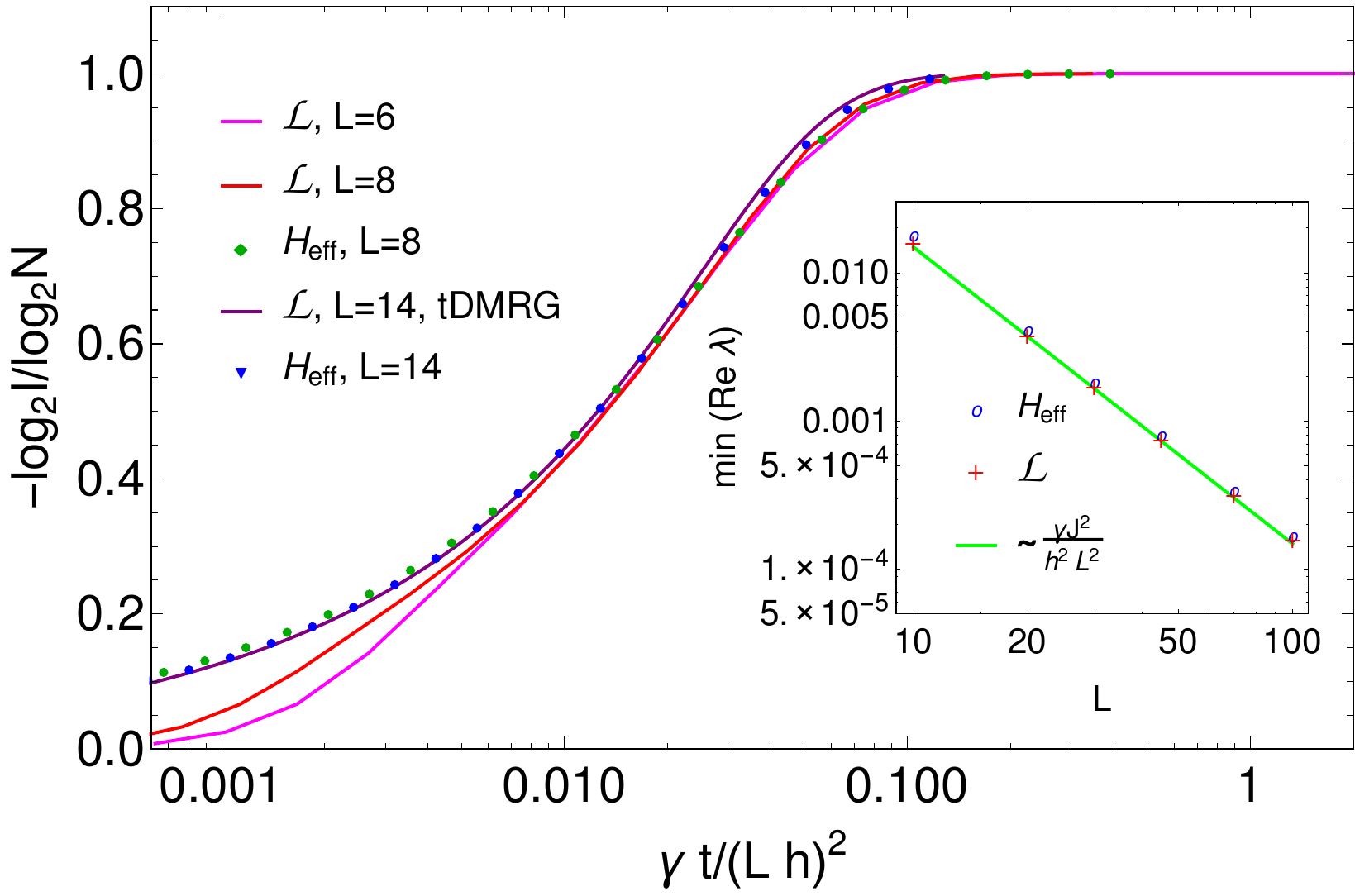}
\vspace{-3mm}
\caption{\label{wall}
(Color online) Scaling function for the domain wall initial state. $N={L \choose L/2}^2$ is the Hilbert space dimension of the sector with zero magnetization~\cite{foot3} ($h=4, \gamma=J=1, \Delta=0.5$). Inset: Scaling of the largest eigenvalue of $\Li$ (the gap) with the system size in the one-magnon sector.
}\end{center}
\end{figure}
While a typical initial state will result in a scaling function that does not depend on $L$, there can be exceptions. As an important example we take a domain wall initial state $\ket{\uparrow \ldots \uparrow\downarrow\ldots \downarrow}$ with zero total magnetization. The effective ferromagnetic $\Heff$, (\ref{eq:Heff}), has a global $SU(2)$ invariance, implying degeneracy. The two lowest eigenmodes (including zero) are the same in all non-trivial magnetization sectors of $\Heff$~\cite{theorem}. Interestingly, even with the disordered bonds the low lying excitations of a ferromagnetic $\Heff$ have a magnon nature. For example, in the one-magnon sector the low-energy (real) eigenvalues of $\Li$ are well approximated by $\lambda_j \sim  \gamma J^2 j^2/(h^2 L^2)$. Numerical demonstration that the spectral gap of many-body ${\cal L}$ as well scales as $\propto 1/L^2$ is shown in the inset of Fig.~\ref{wall}. Low-energy physics of $\Heff$, and with it long time dynamics of $\Li$, is therefore of a hydrodynamic diffusive nature. Because the domain wall initial state has a large overlap with these long wavelength modes (unlike in AF state) the time-dependence of purity gets an additional $L^2$ factor, with the scaling variable being $J^2\gamma t/(h L)^2$, Fig.~\ref{wall}. There is again a good agreement between the full Liouvillian dynamics and the dynamics given by $\Heff$. We note that, since $\tau$ is proportional to the spectral gap of $\Li$, such domain wall initial state could be used to experimentally measure the scaling of the gap~\cite{Gaps} with $L$ -- a spectral quantity of fundamental importance that would be difficult to measure otherwise.

\section{Non-thermodynamic quantity}

\begin{figure}[tb]
\centering \includegraphics[width=3.33in]{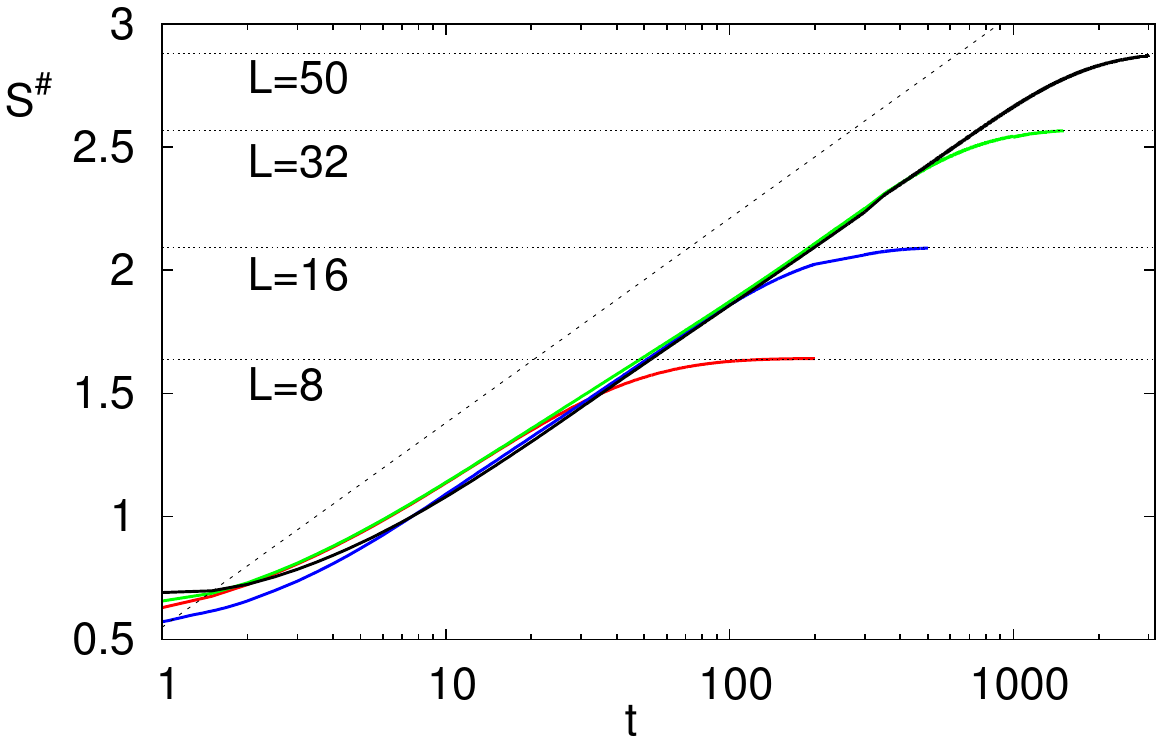}
\vspace{-3mm}
\caption{(Color online) Operator-space entanglement entropy $\opS(t)$ of $\rho(t)$ for the exact evolution with $\Li$ ($\gamma=J=1$, $h=4$, $\Delta=0.5$) and AF initial state. Dotted black line is $\frac{1}{4}\log_2{(t)}+{\rm const.}$, suggesting logarithmic growth. Horizontal dashed lines are the exact theoretical saturation values of $\opS \propto \frac{1}{2}\log_2{L}$ obtained from the steady-state Schmidt coefficients of the projection of $\mathbbm{1}$ to the half-filled subspace~\cite{Gaps}.
}
\label{fig:afoperS}
\end{figure}

So far we have been concerned with $I(t)$, which is a thermodynamic expectation value of $\Heff$, and as such can not signal any possible MBL in $\Heff$. We are now going to consider a quantity which behaves differently. If we make a bipartition of the chain into two halves and Schmidt decompose $\rho(t)$ as $\rho(t)=\sum_k \sqrt{\mu_k} A_k \otimes B_k$, where $A_k$ are orthogonal, $\tr{A_k^\dagger A_p}\propto \delta_{k,p}$, and similarly for $B_k$, and use normalization $\sum_k \mu_k=1$, we can define the operator-space entanglement entropy (OSEE)~\cite{PP07} $\opS:=-\sum_k \mu_k \log_2 \mu_k$. It measures non-factorizability of $\rho(t)$ (it is the entanglement entropy of $\ket{\rho(t)}$ considered as a pure state in the Hilbert space of operators). For MBL systems it has been shown that $\opS(t)$ for initial localized operators grows in the same logarithmic way as $S(t)$ for pure product initial states~\cite{MBL08}. We observe that in our effective description probabilities $
 \vec{p}(t)$ are evolved (\ref{eq:p}) by ${\rm e}^{-\Heff\, t}$ in the same way as would be a pure state in unitary evolution by $\Heff$ (barring the ``imaginary time''). Therefore, the OSEE $\opS(t)$ could behave in a similar way as would von Neumann entropy for evolution of pure states with $\Heff$. We note that $\opS$ is in fact a decisive quantity for the efficiency \cite{PZ07} of our tDMRG simulations of $\rho(t)$, see Fig.~\ref{fig:afoperS}.
One can observe a suggestive logarithmic growth $\opS(t) \sim \frac{1}{4}\log_2(t)$ for over two decades in time, with a prefactor $\frac{1}{4}$ seemingly being equal to $\frac{1}{2\nu}$, where $\nu$ is the critical exponent of the Liouvillian gap, $g \asymp 1/L^\nu$ ($\nu=2$ in our model). Importantly, as opposed to $S(t)$, the OSEE $\opS$ is not extensive, making such logarithmic growth at all possible. Of course, for long times $\opS$ saturates at the steady-state value that scales asymptotically~\cite{Gaps} as $\frac{1}{2}\log_2{[\frac{\pi e}{8}L]}$, and corresponds to the trivial steady state with no MBL.

\section{Conclusion} 

We have shown that the long-time dynamics of the MBL system with dephasing noise is described by an effective stochastic process generated by a bond-disordered ferromagnetic spin chain. Such description becomes more and more precise at large $h$ and is independent in the leading order of $1/h$ from the interaction strength $\Delta$ in the Hamiltonian, explaining experimental observations~\cite{talkBloch}. The scaling variable is shown to be $\gamma t/h^2$ for all observables. The modes which survive at long-times have a hydrodynamic long-range nature. The purity and von Neumann entropy at long times behave as thermodynamic quantities and do not exhibit any MBL signatures. On the other hand, the operator-space entanglement entropy, measuring factorizability in the operator space, does exhibit logarithmic growth in time. Perturbative calculation proposed could be used also for other MBL systems, including some other types of dissipation. We also discuss how, with a right choice of the initial state, one could experimentally ``measure'' the Liouvillian gap.

{\em Note added:} after completion of this work a preprint appeared~\cite{Altman2015} studying the same model, but focusing on different physical properties.

We acknowledge support by grants P1-0044, J1-5439, J1-7279 and N1-0025 of Slovenian Research Agency (ARRS).

\appendix

\section{Deriviation of the effective evolution operator}

In our derivation of the effective evolution operator $\Heff$ we closely follow the Supplementary material of Ref.~\onlinecite{Cai:13}, where a similar derivation has been done, only in the absence of the disorder in the parameters of the model. We also note that in Ref.~\onlinecite{Lesanovsky13} a bit different perturbative method has been used for Rydberg systems, leading in general to similar results. 

The evolution operator of the open system is:
\bea &&\Li [\rho] = -i [H,\rho]+\D[\rho], \nonumber \\
&&H = \sum_{j=1}^{L} J_j\left( \sx{j}\sx{j+1} + \sy{j}\sy{j+1}+\Delta_j \sz{j}\sz{j+1}\right) + \sum_{j=1}^L  h_j\sz{j},\nonumber\\
&&\D[\rho]  = \sum_j \left(2 l_j \rho l_j^\dagger  - \{ l_j^\dagger l_j,\rho\} \right),\quad l_j = \sqrt{\gamma_j/2}\sz{j}, \nonumber 
\eea
Here we put periodic boundary conditions for the Hamiltonian, unlike in the main text, where open boundary conditions are used. We do it in order to have  more symmetric expressions for the couplings of the effective evolution operator. For large systems  one should not expect any difference between periodic and open boundary conditions.

Perturbation expansion is performed with respect to the part of the Liouvillian which is diagonal in the $z$-basis:
\bea 
&& \Li_0 [\rho]= -i [H_{Z},\rho]+\D[\rho], \nonumber \\ 
&& H_Z = \sum_{j=1}^{L} J_j\Delta_j \sz{j}\sz{j+1} + \sum_{j=1}^L  h_j\sz{j}.\eea
For $\Li_0$ the non-equilibrium steady state, $\rho(t=\infty)$, has degeneracy $2^L$ and consists only of diagonal density matrices, $\rho_0^{\kappa} = \ket{\kappa}\bra{\kappa}$, where $\ket{\kappa}\equiv\ket{\psi_n}$ is the $n-$th computation basis vector in the $\sz{j}$-eigenbasis, $\ket{\kappa}=\ket{\sigma_1\sigma_2\ldots\sigma_L}$, $\sz{i}\ket{\kappa}= \sigma_i \ket{\kappa},~\forall i$, and $\sigma_j=1$ for spin up and $\sigma_j=-1$ for spin down. 
The degeneracy is lifted by the hopping between the sites:
\bea 
&& \Li_1 [\rho] = -i [H_{XX},\rho], \nonumber \\
&& H_{XX}=\sum_{j=1}^{L} J_j\left( \sx{j}\sx{j+1} + \sy{j}\sy{j+1} \right). \eea
The action of the $\Li_1$ on the single diagonal density matrix $\rho_0^{\kappa}$ is:
\be \Li_1 \rho_0^{\kappa} =-i \sum_j J_j (\ket{\kappa^{(j)}}\bra{\kappa}-\ket{\kappa}\bra{\kappa^{(j)}}),\ee
where $\ket{\kappa^{(j)}} = \tfrac{1}{2}\left( \sx{j}\sx{j+1} + \sy{j}\sy{j+1} \right)\ket{\kappa}$.
So we see that the first order of the perturbation theory gives zero, as $\langle \kappa \ket{\kappa^{(j)}}=0$, while the second order does not vanish:
\be \Li_{\rm eff} = -\mc{P} \Li_1 (1-\mc{P})\Li_0^{-1}(1-\mc{P})\Li_1 \mc{P}, \label{correction}\ee
where $\mc{P}$ is a projector on the eigenoperators of $\Li_0$:
\be \mc{P}[\rho] = \sum_{\kappa} \tr (\rho \rho_0^{\kappa}) \rho_0^{\kappa}.\ee
$\Li_1$ has non-zero matrix elements only in the subspace orthogonal to $\mc{P}$, therefore the factor $1-\mc{P}$ is not necessary in the expression (\ref{correction}).

We notice that $\ket{\kappa^{(j)}}\ne 0$ if and only if the spins on the neighbouring sites $j,j+1$ point in different directions, $\sigma_j \sigma_{j+1}=-1$.
The eigenvalue of $\Li_0$ with the eigenoperator $\ket{\kappa^{(j)}}\bra{\kappa}$  is
\begin{widetext}
$$\epsilon_{j,\kappa} = 2i \sigma_j \left(h_j - h_{j+1} +\sigma_{j-1}J_{j-1}\Delta_{j-1} - \sigma_{j+2}J_{j+1}\Delta_{j+1} \right) -2 (\gamma_j + \gamma_{j+1})$$ 
and, correspondingly, the one with eigenoperator $\ket{\kappa}\bra{\kappa^{(j)}}$ is
$$\epsilon^\prime_{j,\kappa} = -2i \sigma_j \left(h_j - h_{j+1} +\sigma_{j-1}J_{j-1}\Delta_{j-1} - \sigma_{j+2}J_{j+1}\Delta_{j+1} \right) -2 (\gamma_j + \gamma_{j+1}).$$ 
We notice that $\epsilon$ and $\epsilon^\prime$ are complex conjugated. The eigenvalues $\epsilon$ and $\epsilon^\prime$ depend on the spin projections $\sigma_j$ in $\ket{\kappa}$ at positions $j,~j-1,~j+2$.

Now we can apply $\mc{P}\Li_1$ to $\Li_0^{-1}\Li_1 \rho_0^{\kappa}$:  
\bea -\Li_{\rm eff}\rho_0^{\kappa} = -i\frac{2J_j\mc{P}}{\epsilon_{j,\kappa}}\left[H_{XX} \ket{\kappa^{(j)}}\bra{\kappa} - \ket{\kappa^{(j)}}\bra{\kappa} H_{XX} \right]+i\frac{2 J_j\mc{P}}{\epsilon^*_{j,\kappa}}\left[H_{XX} \ket{\kappa}\bra{\kappa^{(j)}} - \ket{\kappa}\bra{\kappa^{(j)}} H_{XX} \right].
\eea
The very first and the very last term in the expression above give a diagonal contribution:
\be \sum_{j} 2J_j^2\bra{\kappa} (\sz{j}\sz{j+1}-1) \ket{\kappa} \cdot \ket{\kappa}\bra{\kappa} \left(\frac{1}{\epsilon_{j,\kappa}}+\frac{1}{\epsilon^{*}_{j,\kappa} }\right),\ee
while the other two terms change $\ket{\kappa}\bra{\kappa}$ into $\ket{\kappa^{(j)}}\bra{\kappa^{(j)}}$, so they can be written as 
\be  \sum_{j} 2J_j^2 \bra{\kappa^{(j)}} (\sx{j}\sx{j+1}+\sy{j}\sy{j+1}) \ket{\kappa} \cdot \ket{\kappa^{(j)}}\bra{\kappa^{(j)}} \left(\frac{1}{\epsilon_{j,\kappa}}+\frac{1}{\epsilon^{*}_{j,\kappa}} \right).\ee
The multiplier proportional to hoppings $t_{j,\kappa}$ depends only on  $\sigma_{j-1}$ and $\sigma_{j+2}$:
\be t_{j,\kappa} \equiv \frac{1}{\epsilon_{j,\kappa}} + \frac{1}{\epsilon^{*}_{j,\kappa}} = -\frac{(\gamma_j+\gamma_{j+1})}{ \left(h_j - h_{j+1} +\sigma_{j-1}J_{j-1}\Delta_{j-1} - \sigma_{j+2}J_{j+1}\Delta_{j+1} \right)^2 + (\gamma_j+\gamma_{j+1})^2}.\ee
In total the effective evolution operator is: 
\bea \Li_{\rm eff} \ket{\kappa} \bra{\kappa} = \sum_j \sum_{\kappa^{(j)}} 2J_j^2 t_{j,\kappa} \bra{\kappa^{(j)}} (1-\vec{\sigma}_j\cdot\vec{\sigma}_{j+1}) \ket{\kappa} \ket{\kappa^{(j)}} \bra{\kappa^{(j)}}.\eea
Depending on the sign of the spin projection at the positions $j-1$ and $j+2$ there are two possiblities for signs between $J_{j-1}\Delta_{j-1}$ and $J_{j+1}\Delta_{j+1}$: if the spins at the sites $j-1$ and $j+2$ have the same direction, then the sign is minus, while otherwise plus.  Therefore, $\Li_{\rm eff} \ket{\kappa} \bra{\kappa}$ can be also rewritten explicitly as the sum over projectors onto different neighbouring spin-configurations:
\bea
\Li_{\rm eff} \ket{\kappa} \bra{\kappa}  &=& \sum_{\kappa^{\prime}} \bra{\kappa^{\prime}} \Heff \ket{\kappa} \cdot \ket{\kappa^{\prime}} \bra{\kappa^{\prime}}, \\
\Heff &=& \frac{1}{4} \sum_j \sum_{\alpha,\beta=+,-} T_{j,\alpha\beta} (1+\alpha\sz{j-1})   (1- \vec{\sigma}_j\cdot\vec{\sigma}_{j+1} )
(1+\beta \sz{j+2}),~\alpha,\beta=+,-, \label{HEFFfull}\\
T_{j,\alpha\beta} &=& \frac{2J_j^2(\gamma_j+\gamma_{j+1})}{\left(h_j - h_{j+1} + \alpha J_{j-1}\Delta_{j-1} - \beta J_{j+1}\Delta_{j+1} \right)^2 + (\gamma_j+\gamma_{j+1})^2}.
\eea
\end{widetext}
This form underlines the non-local structure of the effective evolution operator. 

The matrix $\Heff$ has the following property:
\be \forall k \in \mathbb{N}: \sum_m \left(\Heff^k \right)_{mn}=0, \quad \sum_n \left(\Heff^k \right)_{mn}=0,\ee 
therefore $\exp(-\Heff t)$ is a doubly stochastic matrix, i.e.  $\sum_m \exp(-\Heff t)_{mn}=1$ as well as $\sum_n \exp(-\Heff t)_{mn}=1$. 

\medskip
{\it Disorder in magnetic field.}
Let us consider a case in which we are interested in the main text of the paper, namely when $h_i \in[-2 h, 2h]$  and $h$ is the largest energy scale of the problem, $\gamma_i,~J_i\equiv J,~J_i\Delta_i \ll h$. Then we could have naively  made a Taylor expansion of the coefficients $T_{j,\alpha\beta}$ with respect to the large parameter $h$. However, this procedure is not well-defined from the mathematical point of view as the difference  $(h_j-h_{j+1})$ is a random variable, with the distribution function non-zero everywhere at the segment $[-4h,4h]$, i.e. there can be also very small values of $(h_j-h_{j+1})$, so the Taylor expansion is not justified. 
In order to have a well-defined mathematical procedure for the large $h$-expansion, we should consider a distribution function for the couplings $T_{j,\alpha\beta}$:
\begin{widetext}
\bea P_{\rm full}\left(Y = T_{j,\alpha\beta} \right) &=& \left[P\left(x=\sqrt{\frac{2 b_j J^2}{Y}-b_j^2}-\mu_j \right)+ P\left(x=-\sqrt{\frac{2 b_j J^2}{Y}-b_j^2}-\mu_j \right) \right] \left(\frac{Y^2}{J^2 b_j} \sqrt{\frac{2 b_j J^2}{Y}-b_j^2} \right)^{-1},\qquad\\
&&P(x) = \begin{cases}
          \frac{4h-x}{4h},~x\geq 0,\\
           \frac{4h+x}{4h},~x< 0,
         \end{cases}
\\
&&b_j=\gamma_j + \gamma_{j+1},\quad \mu_j=\sigma_{j-1}\Delta_{j-1}J-\sigma_{j-2}\Delta_{j+1}J.\eea
\end{widetext}
Compare this with the distribution function for the couplings $T_{j,\alpha\beta}$ which we would have obtained had we performed a naive Taylor expansion:
\be P_{0}\left(Y = \frac{2 b_j J^2}{(h_j-h_{j+1})^2}\right) = J\sqrt{2b_j} \frac{4h\sqrt{Y}-J\sqrt{2b_j}}{(4h)^2 Y^2}.\ee
We see that we obtain $P_0$ from $P_{\rm full}$ for $h\gg\mu_j$ and $\frac{J^2}{Y}\gg b_j$.  The first inequality requires the interaction $\Delta$ to be small comparing to the disorder strength $h$.  The second inequality is rewritten as $Y\ll \frac{J^2}{\gamma}$ as well as we expect that $Y \le \frac{\gamma J^2}{h^2}$, which gives us a condition $\gamma \ll h$ (here we assumed for simplicity $\gamma_j\equiv\gamma$).

The terms containing more than two spins, e.g. $\sz{j-1} (1-\vec{\sigma_j}\vec{\sigma}_{j+1})\sz{j+2}$, are canceled in this order of the expansion (the condition $\Delta \ll h$ is important for such cancelation), so we arrive at $\Heff$ given in the main text.

Let us note that the two distributions,  $P_{\rm full}(Y)$ and $P_0(Y)$, have significantly different behaviour for $Y\rightarrow \infty$: while $P_{\rm full}(Y) = 0$, for $Y>b_j^2/J^2$, $P_0(Y)$ has in infinite tail $P_0(Y) \propto Y^{-3/2}$. However, as we are interested in the long time-dynamics of the propagator $e^{-\Heff t}$, which is driven mostly by the small eigenvalues of $\Heff$, and therefore small $J_{\rm eff}$, the largest eigenvalues of $\Heff$, which are given by the tail of the distribution $P(Y)$,  $Y\rightarrow\infty$, are not important.

%$\gamma \gg (J/16)^{2/3}$ (for constant values of $\gamma$ and $J$; the inequality is obtained combining two inequaties $t> 1/\gamma$ and $J\gg \gamma$) 

%{\it On the long-time dynamics.}
%The long-time dynamics of the system is determined by the weakest links in the effective description. From our expression for the distribution function we might conclude that there is no dynamics beyond $t>1/H^2$ ({\bf check denominator!}), as there are no links in the distribution beyond this edge. But we should remember that $\Heff$ is obtained by the second order perturbation theory. In the next orders of perturbation theory there will be links in $\Heff$ of the magnitude till $1/H^4$ etc. Interestingly these links will be coming with the more and more non-local evolution operator ({\bf check ... if this non-locality will survive the expansion}).  

Let us note that in the one magnon sector, i.e. for $H_{\rm eff}$ restricted to a Hilbert subspace with a single overturned spin, there is no dependence on $\Delta$ at all.

\medskip
{\it Disorder in the interaction strength.} We can also deduce from Eq.~(\ref{HEFFfull}) that at $h_i=0$ and disorder in $\Delta_j$, $\Delta_j\in[-\Delta,\Delta]$, $\Delta\gg 1$, the effective evolution operator is not reduceble to the nearest spin interaction only and contains also four spin terms. 
We notice as well that our effective evolution operator can be used also for describing the dynamics in more than one dimension, as dimensionality of the lattice has not been used in the derivation. 
%Interestingly, the dynamics observed in experiments~\cite{talkBloch} becomes dependent on interaction in the dimension of the space larger then $1$. As we have just seen from the expansion of the probability distribution of the effective couplings, the interaction strength enters in the next order of the expansion with respect to $1/h$. Apparently with increasing the dimension influence of such terms becomes more prominent. Also in higher dimensions the mean-field treatment is more applicable, providing exponential decay for example of imbalance, as the non-local terms are proportional to onsite densities. 

\subsection*{Numerical comparison of the spectrum}
\begin{figure}[t]
\begin{center}
\includegraphics[width=1.03\linewidth]{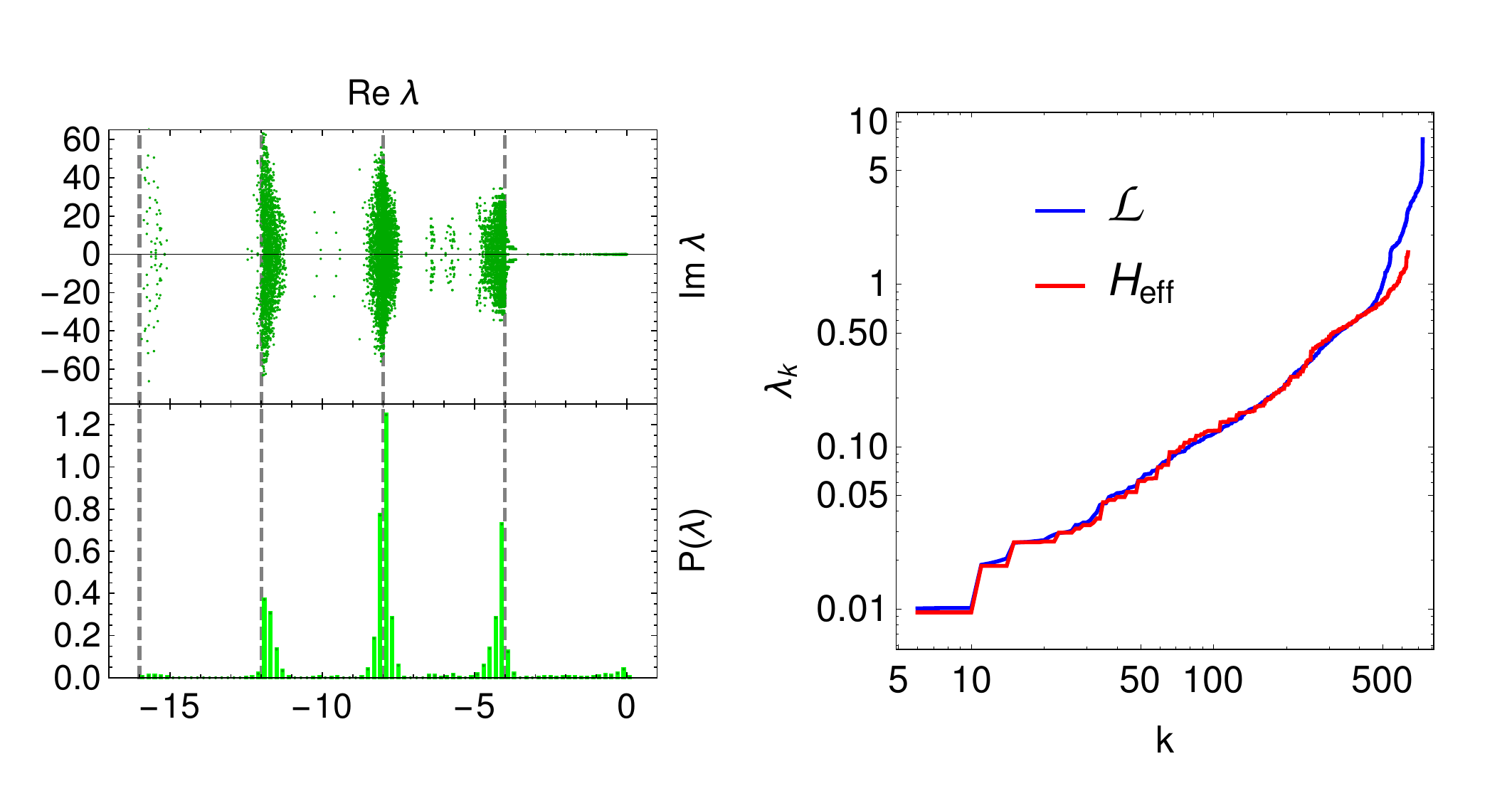}
\vspace{-9mm}
\caption{\label{fig:SpectraMain}
(Color online) Left: The spectrum of the Liouvillian for $L=8$ (in the largest sector where bra and ket have the same magnetization). Lower frame shows the density of real parts of eigenvalues. Right: Comparison of the spectra of $\Heff$ and of purely real eigenvalues of $\Li$ for $L=10$. Parameters are $h=4$, $\gamma=1$, $J=0.5$, $\Delta=1.0$
}\end{center}
\end{figure}

Here we show, see Fig.~\ref{fig:SpectraMain}, the full spectrum of $\Li$ and compare its real eigenvalues  with the spectrum of $\Heff$.
The spectrum of $\Li$ contains $4^L$ eigenvalues. The eigenvalues which determine the long-time dynamics are in clear correspondence with the eigenvalues of $\Heff$ given in the text.   

\section{Averaging purity over the states}

In this section we explicitly show the connection between the purity and the statistical sum of the effective evolution operator. 
The purity is defined as
\be I(t) = \tr \rho^2(t). \ee
Let us express it in terms of the spectral form of $\Heff$: 
\be \Heff = \sum_n \lambda_n \ket{\phi_n}\bra{\phi_n} .\ee
On the other side, as $\Li_{\rm eff}$ acts in the space of the diagonal density matrices, so that the purity is:
\be I(t) = \sum_n p_n^2(t), \ee
where $p_n$ is the $n$th component of the probability vector which evolves as:
\be \ket{p (t)} = \sum_n e^{-\lambda_n t} \ket{\phi_n}  \langle \phi_n | p(0) \rangle.\ee
So it follows that
\begin{widetext}
\be\sum_n p_n^2(t) = \sum_n \left(\sum_{j_1} e^{-\lambda_{j_1} t} \langle \phi_{j_1} | p(0) \rangle \langle \psi_n|\phi_{j_1}\rangle \right)
\left(\sum_{j_2} e^{-\lambda_{j_2} t} \langle \phi_{j_2} | p(0) \rangle \langle \psi_n | \phi_{j_2} \rangle \right).
\ee
\end{widetext}
If we separate $\Heff$ into different magnetization sectors then in each of the sectors eigenvalues are non-degenerate, positive, the operator itself has only real values, so its eigenvectors $\ket{\phi_j}$ have only real entries $\langle \psi_n | \phi_j\rangle$, therefore orthogonality condition between different vectors can be written without complex conjugation:
\be \langle \phi_{j_1} | \phi_{j_2} \rangle  = \delta_{j_1 j_2} \Rightarrow \sum_n \langle\psi_n|\phi_{j_1}\rangle  \langle \psi_n | \phi_{j_2}\rangle  = \delta_{j_1 j_2}.\ee
Hence the purity for the initial density matrix which has only diagonal elements, and therefore is described by the probability vector $\ket{p(0)}$, can be written as
\be I(t) = \sum_j e^{-2\lambda_j t} \mid \langle \phi_j | p(0) \rangle \mid^2.\ee
The purity averaged over initial states in the whole Hilbert space of $\Heff$ is 
\bea && \langle I(t) \rangle_{\mc{H}_{eff}}  = \frac{1}{\mathrm{dim} \Heff } \sum_{j=1}^{\mathrm{dim} \Heff } e^{-2\lambda_j t} \equiv \frac{\tr \exp(-2\Heff t)}{\mathrm{dim} \Heff },  \nonumber \\
&&\mathrm{as}~\langle \mid  \langle \phi_j | p(0) \rangle \mid^2  \rangle_{\Heff} = \frac{1}{\mathrm{dim} \Heff }.
\eea
The dimension of the effective evolution operator $\mathrm{dim} \Heff $ equals to the dimension of the fixed magnetization (particle number) sector under consideration.


\begin{thebibliography}{99}
\bibitem{footg} The effective $\Heff$ (\ref{eq:Heff}) is correct in the limit of large $h$, keeping other parameters ($\Delta$, $\gamma$) fixed. Other limits, for instance, letting $\gamma \to 0$ while keeping other parameters fixed will likely require a different description.
\bibitem{foot2} We note that there is no logarithmic growth in time of $-\log_2 I$ (or $S(t)$) as $f(\tau) \in [0,1]$, whereas the logarithm has the whole positive real axis for its range.
\bibitem{foot1} For a single disorder realization, where the largest bond is finite, for times smaller than this inverse energy, dependence is $-\log I \sim \tau$, leading to $\sim \sqrt{\tau}$ behavior only after disorder averaging.
\bibitem{foot3} In Fig.~\ref{wall} we scale the purity with the exact dimension of the Hilbert space. In the thermodynamic limit the factor $\log_2 N$ and $L-1$ used in other figures coincide.
\end{thebibliography}
\end{document}